\documentclass[prl,twocolumn,showpacs,amsmath,amssymb,superscriptaddress]{revtex4}

\usepackage{graphicx}

\begin{document}
\title{Mesoscopic conductance fluctuations in graphene}

\author{D. W. Horsell}
\author{A. K. Savchenko}
\author{F. V. Tikhonenko}
\affiliation{School of Physics, University of Exeter, Stocker Road, Exeter, EX4 4QL, UK}
\author{K. Kechedzhi}
\affiliation{Department of Physics, Lancaster University, Lancaster, LA1 4YB, UK}
\author{I. V. Lerner}
\affiliation{School of Physics and Astronomy, University of Birmingham, Birmingham B15 2TT, UK}
\author{V. I. Fal'ko}
\affiliation{Department of Physics, Lancaster University, Lancaster, LA1 4YB, UK}


\begin{abstract}
We study fluctuations of the conductance of micron-sized graphene devices as a function of the Fermi energy and magnetic field. The fluctuations are studied in combination with analysis of weak localization which is determined by the same scattering mechanisms. It is shown that the variance of conductance fluctuations depends not only on inelastic scattering that controls dephasing but also on elastic scattering. In particular, contrary to its effect on weak localization, strong intervalley scattering suppresses conductance fluctuations in graphene. The correlation energy, however, is independent of the details of elastic scattering and can be used to determine the electron temperature of graphene structures.
\end{abstract}

\maketitle

\section{Introduction}

Fluctuations of the conductance of electron systems have been studied for many years in metals and electron gases formed in semiconductors \cite{BeenakkerSSP44}. They originate from the interference between phase-coherent electron paths that cross the system,
and in diffusive, phase-coherent systems the conductance fluctuations are universal (UCF): their average amplitude of the order of $e^2/h$ is independent of the mean conductance \cite{LeePRL55}. Another consequence of quantum coherence in diffusive systems is a weak localization (WL) correction to the conductance. It was shown recently that WL in a new two-dimensional system, graphene \cite{NovoselovScience306}, is unusual in that it depends not only on inelastic but also elastic scattering of chiral carriers, both  within and between the two graphene valleys \cite{McCannPRL06,TikhonenkoPRL100}.

As quantum interference lies at the origin of both UCF and WL, it is important to understand if manifestation of UCF in graphene is also different from that in conventional two-dimensional systems. In our earlier studies of WL in monolayer and bilayer graphene \cite{TikhonenkoPRL100, GorbachevPRL98} we have established that quantum interference of carriers within one valley is suppressed, but the presence of significant intervalley scattering makes it clearly detectable in the magnetoconductance. Here we perform the first analysis of conductance fluctuations in graphene, both in the metallic and electro-neutrality (Dirac) regions, which are complemented by WL studies. Using recent theories of UCF in graphene \cite{Kharitonov08010302,Kechedzhi08012394,Kechedzhi08083211},  we show that, similar to WL, the amplitude of the fluctuations can be strongly affected by the intensity of elastic scattering, so that it can be larger than in conventional 2D metals. In graphene flakes, however, significant intervalley scattering reduces it to a value close to that in one-valley 2D systems with non-chiral carriers. In order to detect the unusual properties of UCF in graphene with a significant (up to four times) increase of the variance, one has to fabricate samples with no intravalley suppression of interference as well as weak intervalley scattering.

We also examine the effect of elastic scattering in graphene on the autocorrelation function of the fluctuations as a function of the Fermi energy, and show that under usual experimental conditions the correlation energy is insensitive to the specifics of scattering and can be used as a direct measure of the electron temperature. Methods to determine the true electron temperature become important in graphene-based devices where phonons are poorly coupled to the environment, so that with commonly used currents the overheating of electrons is highly likely. The classical conductivity has a very weak temperature dependence \cite{MorozovPRL100} and therefore is not suitable for this purpose.  Both the WL and the variance of UCF have strong temperature dependences \cite{TikhonenkoPRL100, GorbachevPRL98}, but the magnitudes of these effects depend on the details of inelastic and elastic scattering and so do not readily allow for the extraction of the temperature from them. We show here that analysing the width of the correlation function does unambiguously yield the electron temperature in graphene, as well as in any mesoscopic diffusive conductor.

\section{Theory}

\begin{figure}[!htb]
\includegraphics[width=.7\columnwidth]{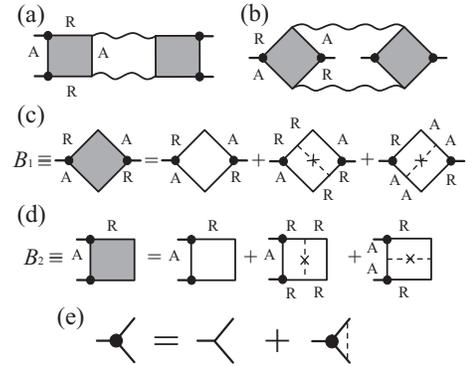}
\caption{(a),(b) The diagrams which contribute to the main order in the diagrammatic expansion of the conductivity-conductivity correlation function. The solid lines are the impurity averaged retarded or advanced Green functions, the short wavy tails are the current vertices and the long wavy lines the diffusion ladders. (c),(d) Hikami boxes of two types and additional diagrams which determine renormalization in the main order in $1/k_{F}l \ll 1$, where the dashed lines correspond to the disorder potential. (e) Diagrammatic equation for renormalized current vertex.}\label{fig:diagrams}
\end{figure}

The variance of the conductance fluctuations in graphene is determined by the standard set of the perturbation theory diagrams in $\hbar/(p_F\ell) \ll 1$ ($\ell$ is the electron mean free path) shown in Fig.~\ref{fig:diagrams}. Structurally, they coincide with the diagrams describing mesoscopic fluctuations in usual conductors~\cite{AltshulerJETPL42} but the Hikami boxes are different \cite{Kechedzhi08012394,Kharitonov08010302,Wurm08081008} because of the linear energy spectrum, carrier chirality and valley degeneracy in graphene. In the low-temperature limit, $L_T\gg L_\varphi$, where $L_T=(\hbar D/k_BT)^{1/2}$ is the thermal diffusion length, $L_\varphi=(D\tau_\varphi)^{1/2}$, ($D$ is the diffusion coefficient and $\tau_\varphi$ is the phase-breaking time), the variance of conductance fluctuations becomes \cite{Kharitonov08010302,Kechedzhi08012394}
	\begin{equation}
	\langle\delta G^2\rangle\equiv\mathcal{F}(0)=
	\dfrac{12}{\beta}g_s^2\mathcal{R}(L,W,L_\varphi,L_i,L_\ast)\, ,
	\label{eqn:ucf}
	\end{equation}
where  $G$ is the conductance in units of $e^2/h$, $\delta G=G-\langle G\rangle$, $W$ and $L$ are, respectively, the sample width and length, $g_s$ is the spin degeneracy, and the time-reversal symmetry parameter $\beta=1$ for $B\ll B_0=(\hbar/e)/L_\varphi^2$ and $\beta=2$ for $B\gg B_0$. The function $\mathcal{R}(L,W,L_\varphi,L_i,L_\ast)$ is given by
	\begin{align}\label{eqn:r}
	\mathcal{R}&=\sum_{n=1,m=0}^{\infty}\,\{\eta_{nm}^{-2}\,+\nonumber\\
	&(\eta_{nm} + 2(L/L_i)^2)^{-2}\,+\\
	2&(\eta_{nm} + ((L/L_i)^2+(L/L_\ast)^2))^{-2}\}\,\nonumber\\
	&=\mathcal{R}_1+\mathcal{R}_2+2\mathcal{R}_3\,\nonumber,
	\end{align}
where $\eta_{nm}=\pi^2n^2+(L/W)^2\pi^2m^2+(L/L_\varphi)^2$, $L_i=(D\tau_i)^{1/2}$ is the intervalley scattering length determined by the scattering rate $\tau_i^{-1}$, and $L_\ast=(D/(\tau_w^{-1}+\tau_z^{-1}))^{1/2}$ stands for the intravalley scattering length determined by both the trigonal-warping scattering rate $\tau_w^{-1}$ and the scattering rate $\tau_z^{-1}$ due to sublattice asymmetric potentials \cite{McCannPRL06}.

In the case of $L_\varphi \gg L$ the conductance variance given by Eqs.~\ref{eqn:ucf} and \ref{eqn:r} is independent of the dephasing length and can be represented as $\mathcal{R}=\alpha\mathcal{R}_1$, $4 \geq \alpha \geq 1$. The coefficient $\alpha$ is sensitive to the strength of intervalley and intravalley scattering in a particular graphene sample. Namely, $\alpha=4$ if $L_i,L_\ast\gg L$; $\alpha=1$ for $L\gg L_i$; and $\alpha \approx 2$ for $L_i>L>L_\ast$. Therefore, the variance of conductance fluctuations in graphene is different from that in a conventional metal \cite{LeePRL55}, in which case $\mathcal{R}=\mathcal{R}_1$ in Eq.~\ref{eqn:r} and the variance is insensitive to the microscopic details of disorder. In the case of a long and narrow graphene sample, strong intervalley scattering at the edges ($L \gg L_i$) results in $\alpha=1$ and $\mathcal{R}_1=\zeta(4)/\pi^4$, leading to $\langle\delta G^2\rangle=(2/\beta)g_s^2/15$, which coincides with the usual result for quasi-1D disordered wires.

At finite temperatures, the autocorrelation function of conductance fluctuations, $F(\Delta)$, is given by a convolution of the ensemble-averaged correlator, $\mathcal{F}(\varepsilon)\equiv\left\langle\delta G(E)\,\delta G(E+\varepsilon)\right\rangle$, and the thermal broadening factor
	\begin{align}
	K(\varepsilon,\Delta) & = \!\int \textrm{d}E
	f'(E,\varepsilon_{\text F}) f'(E + \varepsilon,\varepsilon_{\text
	F}+\Delta)\,,\nonumber
	\end{align}
where  $f'(E,\varepsilon_{\text F}) = -1/(4k_{\text B}T_e) \cosh^{-2}[ (E-\varepsilon_{\text F})/(2k_{\text B}T_e)]$ is the energy derivative of the Fermi--Dirac distribution function ($T_e$ is the electron temperature). Thus
	\begin{align}
	F(\Delta)& \equiv \langle\langle
	\delta G(\varepsilon_{\text F})
	\delta G(\varepsilon_{\text F}+\Delta)\rangle\rangle\nonumber\\
	         & = \int\textrm{d}\varepsilon K(\varepsilon,\Delta)
	         \mathcal{F}(\varepsilon)\,.\label{eqn:finitet}
	\end{align}
Here the brackets $\left\langle \left\langle \cdots \right\rangle \right\rangle $ stand for both the ensemble and thermal averaging and
	\begin{align}
    \mathcal{F} (\varepsilon) &= 4g_s^2 \sum_{n,m}\sum_{l=1,2,3,4}
    \left(\left|\mathcal{D}^l_{nm}\right|^2
    + \frac{1}{2}\mathrm{Re} \left[ \mathcal{D}^{l}_{nm}\right]^2\right),\label{eqn:a-ec}\\
    \mathcal{D}^{1}_{nm} &\equiv \left(-\frac{\textrm{i}}{\hbar} \varepsilon\tau _D
    + \eta_{nm}\right)^{\!-1}\,, \label{eqn:d1}\\
    \mathcal{D}^{2}_{nm} &\equiv \left(-\frac{\textrm{i}}{\hbar} \varepsilon\tau _D
	+ \eta_{nm} + 2(L/L_i)^2\right)^{\!-1}\,, \\
    \mathcal{D}^{3,4}_{nm} &\equiv \left(-\frac{\textrm{i}}{\hbar} \varepsilon\tau _D
	+ \eta_{nm}+ (L/L_i)^2+(L/L_\ast)^2\right)^{\!-1},\label{eqn:d34}
	\end{align}
where $\tau_D\equiv L^2/D$.

The variance of the conductance in graphene and the correlator $\mathcal{F} (\varepsilon)$ given by Eqs.~\ref{eqn:a-ec}-\ref{eqn:d34} are sensitive to the microscopic details of disorder in a particular sample. However, in the normalized autocorrelation function ${F}_\text{n}\equiv F(\Delta)/F(0)$ these details do not manifest themselves under usual experimental conditions.

To show this, let us first consider a narrow wire, $W \ll\min(L_\varphi, L)$, with strong intervalley scattering at the boundaries. Then the part of the correlator $\mathcal{F}$ which contributes to $F_\text{n}$ can be written exclusively via the `valley-singlet' diffusion propagators $\mathcal{D}_{nm}^{1}$. The sum over $m$ in Eq.~\ref{eqn:a-ec} in this case is dominated by the $m=0$ term. (Note that we have taken into account only the diffusion modes and neglected the Cooperons, which corresponds to the regime of suppressed WL by magnetic field.) We obtain an asymptotic expression for $F_n(\Delta)$ assuming $L \gg L_\varphi$ keeping only the term with $m=0$ and performing the summation over $n$:
	\begin{align}\label{int}
    \mathcal{F}(\varepsilon)=4g_s^2 \left(\frac{1}{2\sqrt{2} }
    \left( \frac{L_\varphi }{L_x}  \right)^{\!3}
    \frac{3t^2+t+2}{t^3\sqrt{t+1} }-\left( \frac{L_\varphi }{L_x}  \right)^{\!4}
    \frac{t^2+2}{t^4}\right)\,,
	\end{align}
where $t \equiv\sqrt{(\varepsilon \tau_\varphi /\hbar)^2+1}$. This function is sharply peaked, with a maximum at $\varepsilon=0$ and width $\hbar/\tau_\varphi$. In contrast, the thermal broadening factor $K(\varepsilon,\Delta)$ has a broad peak at $\varepsilon=\Delta$ of width $\sim k_{\text B}T_e \gg \hbar/\tau_{\varphi}$.  Convolution of these functions results in a normalized correlation function that is independent of the microscopic details contained in $\mathcal{F}(\varepsilon )$:
	\begin{gather}
    F_\text{n}(\Delta) = \frac{K(0,\Delta)}{K(0,0)} = \frac{3\left(\theta\coth\theta - 1\right)}{\sinh^{2}\theta}
    ,\;\; \theta\equiv \frac{\Delta}{2k_{\text B}T_e}\,.\label{F}
	\end{gather}
The width of this one-parameter function, determined from $F_n(\theta_c)=0.5$, is $\theta_c=1.36$, which results in
	\begin{align}\label{eqn:main}
    \Delta_c \approx 2.7 k_{\text B}T_e\,.
	\end{align}
This result remains valid under the condition $L_T,L_y\ll \min(L_\varphi, L_x)$ for any $L_\varphi\lesssim L_x$, and allowing for all the diffusion modes in Eqs.~\ref{eqn:d1}-\ref{eqn:d34} in graphene or, indeed, in any other mesoscopic disordered conductor. This result was tested numerically in \cite{Kechedzhi08083211}. There, the values of $\Delta_c$ were found to lie within a narrow interval, $2.7\leq \Delta_c /k_{\text B}T_e\leq 2.9$, within $10\%$ of the asymptotic value $2.7$ of Eq.~\ref{eqn:main}. Good agreement with this value (within $25\%$) was also found when 2D samples were considered.

\section{Experiment}

\begin{figure}[htb]
\includegraphics[width=.7\columnwidth]{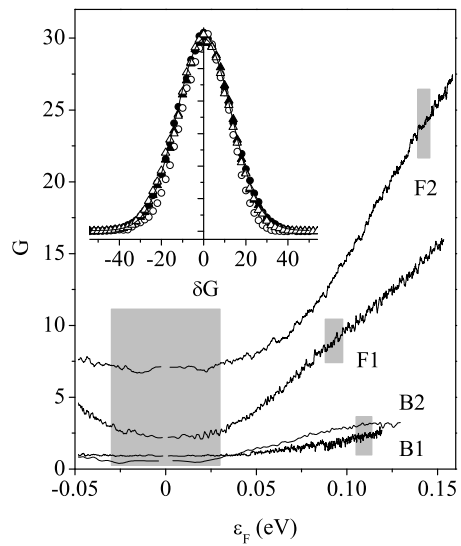}
\caption{Dimensionless conductance of the graphene samples as a function of the Fermi energy at $T=$0.25\,K. (For B2 the hole region is studied so the scale has been reversed to compare with other samples.) The shaded regions show the energy range used in the analysis. The inset shows the distribution function of $\delta G$ for sample F2 averaged over $\sim$14000 `realisations' (filled symbols are for high density, open symbols for the Dirac region).}\label{fig:gef}
\end{figure}

The samples studied experimentally in this work are monolayer graphene flakes created by mechanical exfoliation on a n$^+$Si substrate covered by $300\,$nm of SiO$_2$. In addition to monolayer graphene flakes, we have also studied a bilayer sample (Bi: $L=1.5\,\mathrm{\mu m}$, $W=1.8\,\mathrm{\mu m}$, see \cite{GorbachevPRL98}), to examine the generality of the method of determining the electron temperature. (The number of layers in the studied structures has been established from the analysis of the quantum Hall effect \cite{TikhonenkoPRL100, GorbachevPRL98}.) The manifestation of WL effects in monolayer and bilayer graphene can be very different, but the correlation properties of conductance fluctuations as a function of the Fermi energy are expected to be universal.

The ac current driven through the sample was 1\,nA to avoid overheating (determined by measuring the effect of increasing the current on the mesoscopic fluctuations). The Fermi energy, $\varepsilon_F$, was controlled by a gate voltage, $V_g$, applied between the substrate and the flake (in our monolayers, $\varepsilon_F=\hbar v_F\sqrt{\pi V_g C/e}$, where C is the capacitance per unit area between the gate electrode and graphene). The conductance of all samples is shown as a function of Fermi energy in Fig.~\ref{fig:gef}. The dephasing rate was determined from a fit of the magnetic field dependence of the sample conductance to the theory of weak localization in graphene \cite{McCannPRL06,TikhonenkoPRL100}. (Indeed, in order to analyse the UCF properly the values of not only $L_\varphi$ but also $L_i$ and $L_\ast$ need to first be determined and this can only be done in combination with analysis of the WL.) The sample parameters are given in \cite{Kechedzhi08083211} and extra details of samples F1 and F2 are given in Table~\ref{tab:samples}. Samples B1 and B2 are quasi-1D samples with width $W=0.3\,\mathrm{\mu m}$  and length $L=3.7$ and $2.0\,\mathrm{\mu m}$, respectively.

\begin{table}
\begin{tabular}{c|cccccccc}
Sample & $L$ & $W$ & $L_\varphi^{0.25\,\mathrm{K}}$ & $L_\varphi^{4\,\mathrm{K}}$ & $L_i$ & $L_\ast$ & $\ell$ & $n$ \\
\hline
F1 & 4.1 & 1.8 & 1.5 & 0.7 & 0.4  & 0.06 & 0.07 & 1.4  \\
F2 & 3.8 & 1.8 & 3.8 & 1.7 & 0.5  & 0.1  & 0.1  & 0.7
\end{tabular}
\caption{Lengthscales (in microns) of samples F1 and F2 at high carrier density. The values of the dephasing length, $L_\varphi$ have been determined from analysis of the weak localization. Shown here are values at $T=0.25\,$K ($L_\varphi^{0.25\,\mathrm{K}}$) and $4\,$K ($L_\varphi^{4\,\mathrm{K}}$). $\ell$ is the mean free path and $n$ is the carrier density in $10^{12}$\,cm$^{-2}$.}\label{tab:samples}
\end{table}

Fluctuations of the conductance occur as a function of both the Fermi energy and perpendicular magnetic field. The fingerprint of the fluctuations is robust and temperature dependent over the whole experimental range of 0.26 to 20\,K. For a particular range of energies and fields the autocorrelation function $F(\Delta)$ of the fingerprints can be analysed to determine the variance $F(0)$ and the correlation energy $\Delta_c$ from $F(\Delta_c)=F(0)/2$. The $V_g$ range was chosen to incorporate a sufficient number ($\sim100$) of fluctuations without significantly changing the average value of the conductance. In our experiments this limits the range of temperatures to $T<10\,$K. To increase the number of fluctuations (sample `realisations') the measurements were performed at several ($10-20$) values of magnetic field in the range $B\gg B_0$.

\begin{figure}[htb]
\includegraphics[width=.9\columnwidth]{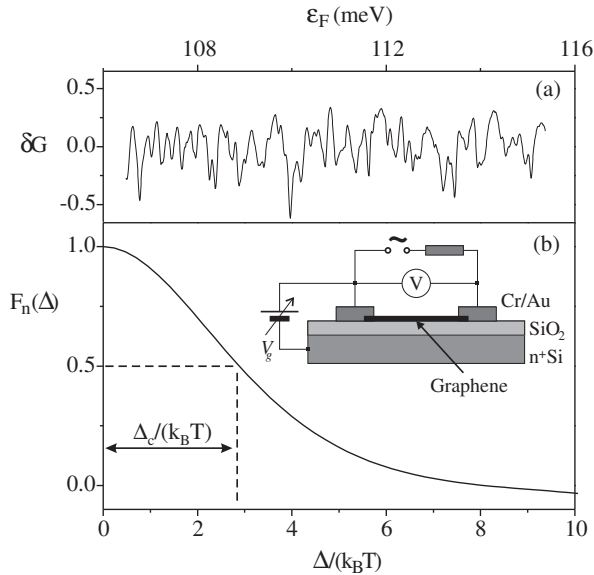}
\caption{(a) Typical fingerprint of $\delta G$ (normalized by $e^2/h$) for sample B1 at $T=0.25$\,K and $B=90$\,mT as a function of Fermi energy. (b) Normalized correlation function for sample B1. The inset shows the circuit used in the experiment and construction of the graphene sample.}\label{fig:acf}
\end{figure}

Figure~\ref{fig:acf}(a) shows a typical fingerprint of conductance fluctuations (with average background, $\langle G\rangle$, removed) in sample B1 as a function of the Fermi energy in a magnetic field of 90\,mT. In Fig.~\ref{fig:acf}(b) the (normalized) autocorrelation function of the fingerprint in (a) is shown. In the inset of the figure the circuit used in the measurements is shown.


\begin{figure}[!htb]
\includegraphics[width=.8\columnwidth]{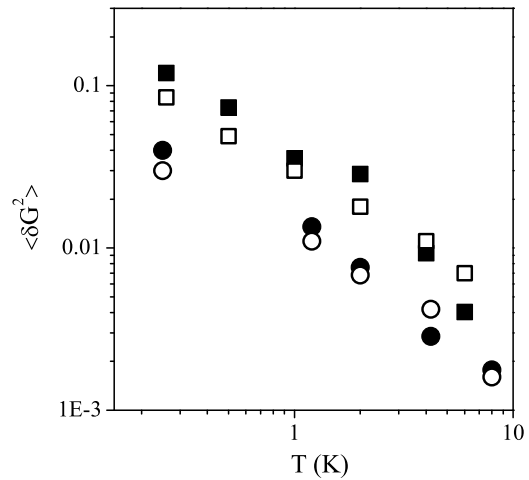}
\caption{The variance of the conductance fluctuations of sample F1 (circles)  with those of sample F2 (squares). Solid symbols correspond to the experimental results  while open ones represent theoretical values calculated using  Eqs.~\ref{eqn:finitet}-\ref{eqn:d34} for $F(0)$.}\label{fig:f1f2var}
\end{figure}

The random nature of the fluctuations is demonstrated for sample F2 in the inset of Fig.~\ref{fig:gef} where the distribution of the conductance fluctuations is seen to have a Gaussian shape typical of UCF \cite{TsyplyatyevPRB68}. (It is also seen that for this sample the magnitude of the fluctuations is very similar at high carrier density and in the Dirac region.) Figure~\ref{fig:f1f2var} shows the variance of samples F1 and F2 at high carrier density. One can see immediately that the variance is temperature dependent over the whole range of temperatures down to 0.25\,K. In the WL experiment, however, it was found that the magnitude of the quantum correction determined by $L_\varphi$ saturates at low temperatures \cite{TikhonenkoPRL100} (when $L_\varphi\sim L$). This highlights the importance of the lengthscale $L_T$ for the amplitude of UCF in 2D, when $L_T<L_\varphi$  \cite{LSFPRB87}.

\section{Analysis and discussion}

Using the values of $L_\varphi$, $L_i$ and $L_\ast$ determined from weak localization measurements in Eq.~\ref{eqn:finitet} we are able to calculate numerically the theoretical values of the variance for our samples. For F1 and F2, $L_\varphi\sim L>L_i\gg L_\ast$. One can see from Eq.~\ref{eqn:r} that the last two terms in $\mathcal{R}$ are suppressed by the strong elastic scattering and therefore $\alpha\sim1$. The good agreement between theory and experiment, Fig. 4, shows that $L_\varphi$ calculated from WL and UCF is the same quantity, as predicted in \cite{AleinerPRB65}, and also that unlike its effect in WL intervalley scattering acts to suppress the conductance fluctuations. In \cite{TikhonenkoPRL100} it was shown that there is strong intravalley suppression of quantum interference in graphene deposited on silica, hence the four-fold increase of conductance fluctuations (Theory section) will not be seen. In addition, due to intervalley scattering at the sample edges the intervalley diffusion length $L_i$ has an upper limit of the sample width. Therefore in narrow graphene samples it is difficult to achieve conditions where $\alpha>1$, and as a result the amplitude of fluctuations is similar to that observed in a usual diffusive metal. We found previously \cite{GorbachevPhysicaE40} that the fluctuations in bilayer sample Bi can also be interpreted using a standard UCF model. This shows a dramatic contrast with WL where in the presence of intervalley scattering the shape of the magnetoconductance curve shows strong dependence on small variations in the elastic scattering.


\begin{figure}[htb]
\includegraphics[width=.9\columnwidth]{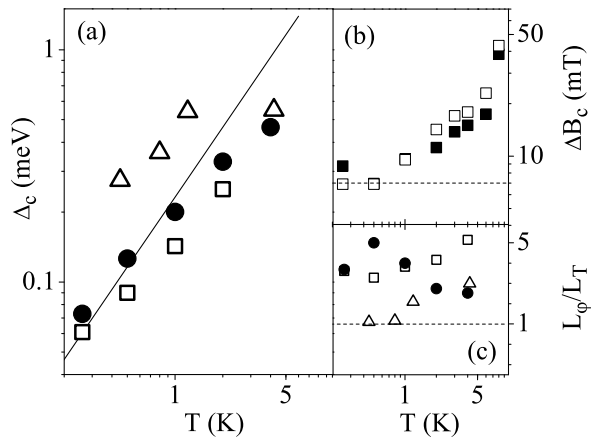}
\caption{(a) Experimental value of the correlation function width $\Delta_c$ for various bath temperatures extracted from the UCF in four graphene-based devices at high carrier density (F2: open squares, B1: filled circles, Bi: open triangles). The solid line corresponds to Eq.~\ref{eqn:main}. (b) Correlation magnetic field of sample B2. Filled squares are experimental results and open squares are $\Delta B_c=(h/e)/WL_\varphi$. The dashed horizontal line shows $\Delta B_c=(h/e)/WL$. (c) Ratio $L_\varphi$/$L_T$ for high carrier densities in samples F2, B1 and Bi.}\label{fig:delta}
\end{figure}

The values of the correlation energy extracted from the correlation functions are shown in Fig.~\ref{fig:delta}(a). It can be seen that the correlation energy increases linearly with increasing bath temperature. Equation~\ref{eqn:main} is plotted in Fig.~\ref{fig:delta} as a solid line. There is agreement between the theory and experiment for all the samples. This includes the bilayer sample (open triangles in figure), which contrasts with the significant difference between monolayer and bilayer samples in the manifestation of weak localization \cite{KechedzhiPRL98,GorbachevPRL98,HorsellPTRSA366}. This also indicates general applicability of the developed model to all diffusive systems. The criterion of the applicability of the theoretical result in Eq.~\ref{eqn:main},  $L_\varphi$/$L_T>1$, is seen to be satisfied in the studied samples, Fig.~\ref{fig:delta}(c).

We want to note that the results of our experiments on graphene show some general properties of UCF \cite{LeePRL55}. Increasing magnetic field above $B_0$ has shown the conventional decrease of the variance by a factor of two. Also, the correlation magnetic field has been seen to be related to $L_\varphi(T)$ in the usual way. Shown in Fig.~\ref{fig:delta}(b) is the correlation field for sample B2 as a function of temperature. For this sample $L_\varphi>W$, so the expected dependence of the correlation field is $\Delta B_c(T)\approx (h/e)/WL_\varphi(T)$ \cite{BeenakkerSSP44}. This is plotted as open squares in the figure and clear agreement is seen with experiment, including the saturation of $\Delta B_c$ that occurs at low temperatures when the dephasing length approaches the sample length.


Finally we discuss the results of our measurements in the Dirac region where the applicability of the developed analytical theory of UCF is not obvious, although the conductivity of our samples satisfies the commonly used criterion for the diffusion theory, $\langle G\rangle L/W\gg 1$.  (We have also shown previously \cite{TikhonenkoPRL100} that the diffusive theory of weak localization describes well the magnetoconductance in the Dirac region.) Numerical investigation of the conductance fluctuations in graphene (from sample to sample) \cite{BeenakkerEL07} have shown that their amplitude is considerably stronger than in conventional metals. At the same time experiments on few-layer samples \cite{Staley07090493} has claimed that conductance fluctuations (as a function of the carrier density) are suppressed near the charge neutrality point.

\begin{figure}[!htb]
\includegraphics[width=.7\columnwidth]{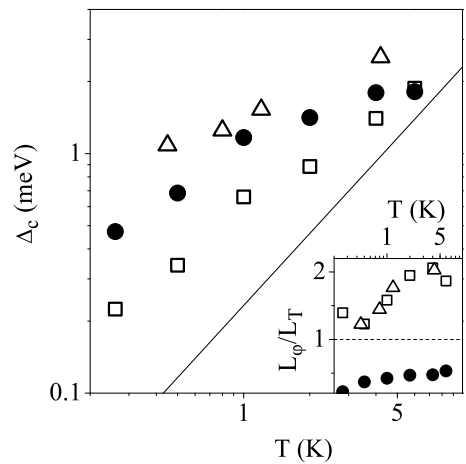}
\caption{$\Delta_c$ as a function of bath temperature at carrier density close to the Dirac point (F2: open squares, B1: filled circles, Bi: open triangles). The solid line corresponds to Eq.~\ref{eqn:main}. Inset: Ratio $L_\varphi$/$L_T$ at low carrier density in samples F2, B1 and Bi.}\label{fig:dirac}
\end{figure}

In our experiments we consider a region of carrier densities around the Dirac point, $-30<\varepsilon_F<30\,$meV. Firstly, we do not observe an increase of the amplitude of conductance fluctuations in the Dirac region compared with that in the high-density regions. In samples F1 and F2 the variance in the Dirac region is very close to that in high-density regions, while in narrow samples B1 and B2 it is several times smaller. Secondly, it can be seen from Fig.~\ref{fig:dirac} that the correlation energy deviates from the `high temperature' limit at low temperatures. This can be attributed to the fact that for all samples the high-temperature condition is now destroyed because of a decrease of the dephasing length by $>30\%$ that occurs when moving from high to low carrier density, Fig.~\ref{fig:dirac}(inset). (This decrease of the low-temperature dephasing length in the Dirac region was obtained from the analysis of WL in both monolayer and bilayer graphene samples \cite{TikhonenkoPRL100, GorbachevPRL98}.)

\section{Conclusion}
In conclusion, in combined studies of conductance fluctuations and weak localization in monolayer and bilayer flakes we have demonstrated that the variance of universal conductance fluctuations in graphene is strongly affected by elastic scattering, in particular intervalley scattering. However, the correlation energy of the fluctuations as a function of the Fermi energy is insensitive to these scattering mechanisms under common experimental conditions. Analysis of this correlation energy allows direct measurement of the electron temperature in graphene structures. We have also discussed the evolution of the variance and the correlation energy of the conductance fluctuations from the region of high carrier density to the region close to the point of electro-neutrality.

\section{}

We gratefully acknowledge financial support from the EPSRC (EP/D031109), the Lancaster--EPSRC Portfolio Partnership (EP/C511743) and ESF FoNE CRP `SpiCo'. We also thank R.~V.~Gorbachev for graphene sample preparation and P.~R.~Wilkins for technical support.

\end{document}